# Ageing Mitigation and Loss Control Through Ripple Management in Dynamically Reconfigurable Batteries

T. Kacetl, J. Kacetl, N. Tashakor, and S. M. Goetz

*Abstract*—Dynamically reconfigurable batteries merge battery management with output formation in ac and dc batteries, increasing the available charge, power, and life time. However, the combined ripple generated by the load and the internal reconfiguration can degrade the battery. This paper introduces that the frequency range of the ripple matters for degradation and loss. It presents a novel control method that reduces the low-frequency ripple of dynamically reconfigurable battery technology to reduce cell ageing and loss. It furthermore shifts the residual ripple to higher frequencies where the lower impedance reduces heating and the dielectric capacitance of electrodes and electrolyte shunt the current around the electrochemical reactions.

## I. INTRODUCTION

### A. Reconfigurable batteries

Electromobility and grid storage are rapidly developing applications of power electronics and batteries. They use battery packs as an energy tank and a semiconductor inverter to generate the ac output for the motor or grid. Conventionally, cells are hard-wired in a battery pack with certain fixed parallel and serial configuration[1]. In combination with an inverter, the ac side of the inverter supplies the grid or an electric motor with ac current, whereas the dc link of the inverter loads the battery with a current resulting from the operation of the inverter[2, 3].

Alternatively, modular circuit structures such as modular multilevel converters (MMC) or cascaded H bridges (CHB) with batteries offer interesting advantages and can form battery systems with immediate multiphase ac output [3, 4]. In contrast to hard-wired batteries, the distributed power electronics can dynamically reconfigure the module interconnection and control the power of individual modules. Thus, reconfigurable MMC–battery systems offer excellent balancing of the state of charge [5, 6] and state of health [7-10], introduce fault tolerance by bypassing defective modules or even semiconductors [11-14], and increase the effectively available capacitance of battery systems [6, 15, 16]. Several companies are developing or already market commercial systems based on battery-integrated CHB/MMC [17-19]. Further advantages over and comparison of reconfigurable with hard-wired batteries can be found in the literature [20, 21].

### B. Module load-ripple

In all CHB circuits, the load currents of the individual module batteries depend on the macro-level topology. In case of a star configuration and ac output, the modules are divided into phase strings, where each of the phase strings supplies one output phase, for instance feeding a motor or the grid. As a result, the module load is rippled, where the spectrum of the module load contains a strong 2nd harmonic of the output ac frequency [22-24]. CHBs without parallel module connectivity alternate between a series module state, where modules run on phase load, and by-pass state, where modules have zero load [25]. Alternating these states introduces components of variable frequency in the module load spectrum. Thus, they depend on the specific control strategy. Such ripple load on the battery cells occurs additionally to the load current and does not contribute to the active output power but constitutes reactive power fluctuations. As the reactive ripple current loads the equivalent resistances of module components and connections, it generates unnecessary additional loss as well as heating. Such extra heating can easily cause derating in thermally limited automotive batteries, and may further degrade components.

Among the various CHB topologies, those with parallel connectivity (e.g., CHB2 in Figure 1, sometimes also denoted as modular multilevel series parallel converters, MMSPC) [25-28] offer better load distribution among battery modules, lower effective source impedance, and lower ripple load, which is a major advantage particularly for battery applications [29]. In CHB² or MMSPC, the parallel states may substitute the inactive bypass state. Paralleling modules eliminates no-load states and lowers the load of the active modules to bring both closer to the mean [30, 31].

In addition to the averaging effect of two parallel modules within a phase string instead of by-passing one, MMSPC topologies allow a reduction of the module load ripple through internal compensation currents and voltages as well as a double neutral point (see Figure 1 at the right end), which allows module paralleling across the previously widely independent phase module strings [22, 29-34]. Dynamical module paralleling across the module strings through this double neutral point can exchange power and improve the load distribution.

### C. High-frequency-shunting dielectric capacitance

During operation, the elements and materials of the battery cells in the reconfigurable battery undergo degradation processes so that the cells gradually lose their capacity and performance [35-40]. The degradation is a result of many physi-cal and particularly electrochemical processes, also called faradaic processes. Rippled load, however, does not necessarily lead to faradaic processes. Electrically charged electrodes and the adjacent electrolyte form a charge double layer [41-44]. Due to the small spacing of relatively large charges, the capacitance of the double layer can be immense. Furthermore, the electrolytes of modern batteries are strongly polar for high lithium-ion solubility, entailing a large dielectric constant as a side product [45]. Thus, the resulting merely dielectric capacitance can absorb enough charge for short current pulses without further chemical reactions. In contrast to the electrochemical capacitance, the dielectric capacitance offers lower impedance and cell heating.

Electrochemical reactions have limited kinetics, often constrained by diffusion, and electrically appear as low-pass system. To initiate sufficient chemical reactions, a high current has to be maintained for an extended period of time in the range of milliseconds [46]. Electrochemical models show that short pulses are almost completely buffered by the dielectric capacitance of the electrodes [47].

The actual contribution of the rippled load to ageing is lately receiving more attention, and experiments agree with the degradation-neutral dielectric shunting at higher frequencies [48]. Measurements suggest the existence of a corner frequency or transition band above which the dielectric charge absorption capability of electrodes dominates and leads to a decrease in ripple-related battery ageing [49]. Operation in this transition band is therefore accompanied by a de-crease in the cell impedance for the battery ripple.

This paper for the first time solves a major problem of MMCs with batteries and other reconfigurable battery systems, specifically their large low-frequency ripple load, which particularly arises in realistic real-world setups that use bandwidth-limited sensors, control busses, and/or control hardware that introduces latency. The control approach aims to exploit the filtering effect of the dielectric electrode capacitance. The method introduces a battery ripple modulation loop in the scheduling algorithm, which shifts the load ripple toward higher frequencies. Considering practical limitations of the monitoring and communication system speed, the control meth-od uses state observers to sufficiently increase the bandwidth of the battery ripple modulation loop. According to aforementioned studies, load content in the high frequency range is absorbed by the electrode capacitance, which significantly lowers the ageing potential driven by electrochemical reactions. Utilization of the electrode capacitance is furthermore associated with lower impedance and consequently losses [50].

## II. CONTROL IN BATTERY-INTEGRATED CHB

CHB topologies, including CHB², incorporate low-voltage semiconductor switches into each module. The high number of individually governable active components provides the degrees of freedom to control additional objectives, such as active balancing of the module charge [51-56]. While online-optimizing model-predictive approaches can trade off all degrees of freedom but suffer from the high computational load, phase-shifted carrier control, for instance, is a rather simple way to manage the complexity and can further use the parallel mode to maximize utilization of the modules [27, 31-33, 57-59]. The majority of the control methods introduces and considers only certain useful switch states on the module level to reduce the complexity, such as parallel P, series plus S+, bypass plus B+, series minus S–, and bypass minus B– [26]. The complexity can be further reduced by introducing feasible series–parallel configurations of the whole string of modules, so called string states S.

A recently presented approach for controlling those expands above multi-objective optimization approaches of module states and includes criteria such as SoC, temperature T, phase current demand $i^*$, measured phase current $i_m$, and the previous string state $S^{-1}$ for the selection of the next string state to comply with the voltage demand $v^*$, discretized in the voltage modulator to $v_d^*$ [60]. However, this method as other solutions with already product-ready topologies with a bus system to distribute the commands, decouple the battery control objectives from the output control to deal with the limited bus capacity. Instead a fast, strictly real-time loop of the controller selects optimal module-string states $S_o$ from an optimized state list provided by a slower loop with more time (see Figure 2). Update times on the second level are typical, but introduces persistent and regular switching patterns that get translated to low-frequency and even sub-harmonic ripple content in the module load. These low-frequency patterns are a result of the interaction of various control objectives and the reality of limited control bandwidth and feedback speed.

Increasing the update rate is limited by relying on slow ac-quisition of module information, which is often collected through data communication busses with considerable latency [61-65]. The communication bus latency in addition to signal processing can readily reach 10 – 100 ms, while the switching rate of the phase voltage period is in the microsecond range. The other existing solutions struggle (and fail) in view of one fundamental trade-off: the need for lab-grade low-latency sensors, fast direct, bus-less connection of all sensors as well as gate signals, and high-performance embedded control for rapid scheduling as described, for instance, in Li et al. [58] clashes with the conditions in more realistic larger systems as used in commercial setups. In commercial systems, a larger number of modules is typically connected to a more economic off-the-shelf controller via a communication bus, more affordable industry-grade sensors provide slower and lower-bandwidth data, and off-the-shelf economic processing power introduces bottle necks as described in Specht et al. [60], Rietmann et al. [66], and Hao et al. [62].

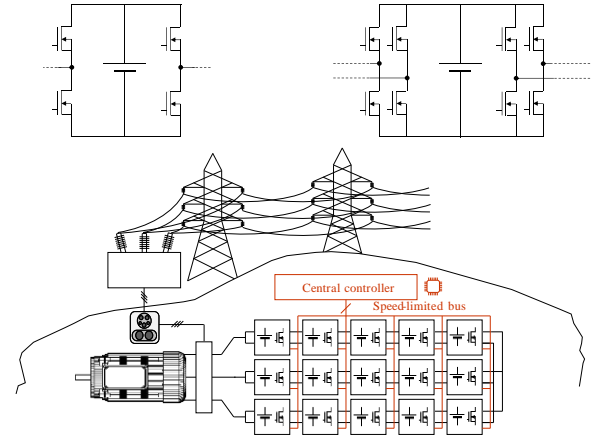

**Figure 1.** Top: Diagram of the system topology: battery module with CHB (left) and CHB2 (right) switch topology. Bottom: Overall system topology in drive trains or with grid connection as storage or charging vehicle.

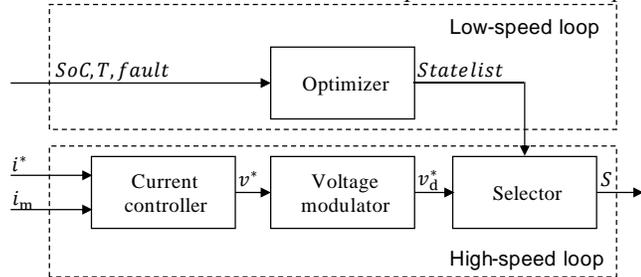

**Figure 2.** Parallel asynchronous optimization for MMC control according to the state of the art, where a low-speed loop optimizes module states or state transitions and stores them in a list or look-up table, which a high-speed loop of the actual controller, modulator, and scheduler uses to actuate the transistors in the modules. Parallel asynchronous optimization methods substantially reduced the computational burden for online optimization.

## III. PROPOSED CONTROL APPROACH

To achieve optimal battery treatment, the load distribution and the ripple modulation become the major, constrained by the voltage demand of the modulator. We use two major components, a strictly real-time compliant state selection as well as the high-bandwidth but asynchronous and not strictly real-time state optimization. The latter writes information into look-up table, which the former reads, enabling concurrent operation. The state optimizer comprises a set of ripple modulators, one assigned to each module, an optimization routine for the selection of optimal module-string states, and a battery-ripple observer, which closes the control loop. The battery-ripple observer instead of waiting for slow measurement data is the core element that enables the high bandwidth. Figure 3 outlines the complete control loop.

Our algorithm aims for a maximization of the bandwidth, which reduces artefactual patterns in switching and reduces loss as well as battery ageing potential associated with low-frequency load ripple. Efficient suppression of switching patterns follows from matching the module control loop and the phase control loop in speed. Therefore, in contrast to previous suggestions in the literature, all blocks of the ripple modulator preferably run within a switching period of the phase control loop. The ripple modulator loop still has a fundamentally asynchronous design and is not strictly time-critical so that any delay in execution does not halt control but might only introduces a short artefactual switching pattern and small ripple with length respective to the delay.

### A. Battery-ripple modulator

The battery-ripple modulator guarantees discharging and charging of the modules at the demanded rate $I^*$ and provides an interface for any higher-level entity that balances the SoC and for any BMS functionality. The use of proportional units in the demanded discharge rate reference $I^*$ distributing the load between the modules ignores one degree of freedom, i.e., scaling, and solves eventual contradictions with the phase-current demand $i^*$.

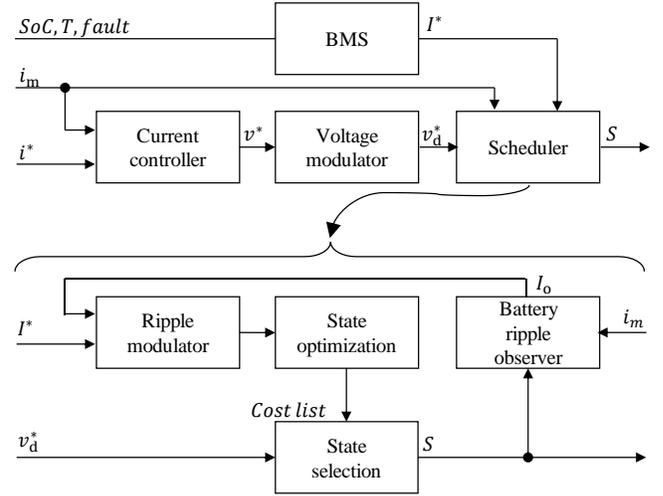

**Figure 3.** Block diagram of the proposed MMC control algorithm with fast quasi-inline optimization inside the scheduler block, which is enabled through a fast module current observer and further detailed at the bottom.

The modulator needs to implement a controller with highly integrational character, also known as reset controller, which brings a controlled variable close to the demanded value. The requirement follows from the distinct distribution $J$ of the phase load $i_\text{m}$ among modules in each string state $S$, which does not necessarily allow the demanded current distribution in each step. The integrated value of the battery-ripple is passed to the optimization routine, which modulates an appropriate sequence of string states and provides the demand on average while controlling the battery ripple.

### B. Battery-ripple observer

The ripple-modulator loop requires considerably fast feedback to run at maximum speed. An acquisition of the module current and transmission of the measured value to the controller represents either unacceptable propagation delay and/or heavy load of a data bus [60, 62, 65-67]. The typical period of the data acquisition is on the order of milliseconds, which is comparable to the load frequency and may not be sufficient for proper control of the module load frequency. Our modulator architecture solves the slacking feedback using an observation technique. The observation technique needs to sufficiently approximate the module load but primarily provide minimal delay. We further derive a simplified observation technique with afore-mentioned qualities.

The actual current load $i_\text{bi}$ of module $i$ is a result of the phase string state (series–parallel configuration of modules), current load of the phase $I_\text{L}$, and voltage $V_\text{B}$ as well as impedance ratios of modules $R_\text{B}$ vs. their interconnection paths $R_\text{S}$ ($R_\text{LS}$ designates resistance of the low-side and $R_\text{HS}$ designates resistance of the high-side interconnection). In principle, the string configuration comprises a set of parallel groups, where each parallel group forms one level of the output voltage and is loaded by the phase current. Equation set

$$\begin{bmatrix} R_{\text{D}1} & R_{\text{U}2} & 0 & 0 & 0 \\ R_{\text{L}1} & R_{\text{D}2} & R_{\text{U}3} & 0 & 0 \\ R_{\text{L}2} & R_{\text{L}2} & R_{\text{D}3} & R_{\text{U}4} & 0 \\ R_{\text{L}3} & R_{\text{L}3} & R_{\text{L}3} & R_{\text{D}4} & R_{\text{U}5} \\ 1 & 1 & 1 & 1 & 1 \end{bmatrix} \cdot \begin{bmatrix} i_{\text{b}1} \\ i_{\text{b}2} \\ i_{\text{b}3} \\ i_{\text{b}4} \\ i_{\text{b}5} \end{bmatrix} = \begin{bmatrix} B_1 \\ B_2 \\ B_3 \\ B_4 \\ I_\text{L} \end{bmatrix},$$

where

$$R_{Dj} = -(R_{Bj} + R_{LSj} + R_{HSj})$$
$$R_{Lj} = -(R_{HSj} + R_{LSj})$$
$$R_{Uj} = R_{Bj}$$
$$B_j = V_{Bj+1} - V_{Bj} - R_{LSj} \cdot I_L$$

(2)

governs further distribution of the phase current among paralleled modules, where the dimension of the problem is equal to the number of parallel modules, and subscript j designates the index of the module in the phase string [68].

Considering constant values of all resistances allows pre-calculation of the impedance matrix and significantly simplifies the algorithm complexity solving (1). The current distribution is then reduced to a function of module voltage differences, which are kept minimal and change only slowly. Under the assumption of relatively constant module voltages within the relevant periods, the whole problem can be further simplified to a look-up table where the observer pre-estimates the expected module current distribution of each feasible string state in the look-up table.

### C. State optimization

The optimization routine also deals with the special setting in case of additional parallel connectivity. To equally distribute the phase load, the majority of modules preferably stay in the active state and rather control their contribution to the phase current by appropriate clustering in parallel groups. Similar to the battery-ripple observer, all string states in the optimization block are represented by the current distribution (in a look-up table). The optimization routine selects output state $S_o$ with optimal current distribution $J_{i,m}$ respecting the demand of the module current controllers $J_i^*$. Our algorithm uses a typical least-square criterion of optimality to evaluate each state. The least-squares criterion guarantees sufficient effort to meet the regulator demands and simultaneously reduces conduction losses by preventing states with far outlying load distribution (e.g., excessive use of bypassing). Results of the evaluation of each feasible state are stored in a look-up table, which interfaces and decouples the ripple modulator loop and the strictly real-time phase-current control loop as outlined above (see Figure 4).

To achieve a sufficient speed of the optimization routine, we constrain transitions between consecutive string configurations by limiting the number of switches that can toggle in each step. This rule also limits the number of commutating switches and consequently reduces switching losses.

### IV. Experimental Ripple Suppression Measurements

We built a single-phase laboratory test setup for experimental evaluation and prepared two control methods: the presented one and as a reference from the state of the art the method of [60]. The aim of the demonstration is to illustrate the effect of the fast feedback and control loop in preventing fixed switching patterns for long intervals.

The test setup includes five CHB² modules with silicon field-effect transistors (IAUT300N10S5N015, Infineon). The dc bus of each module contains a six-cell LiFePO4 battery (22.5 V, 6s, 6.2 Ah). The system controller uses a Mars ZX3 module (Enclustra) with Xilinx's Zynq-7020 system-on-chip. The control algorithm runs fully on the FPGA part. The setup implements a sigma–delta modulator running at 20 kHz.

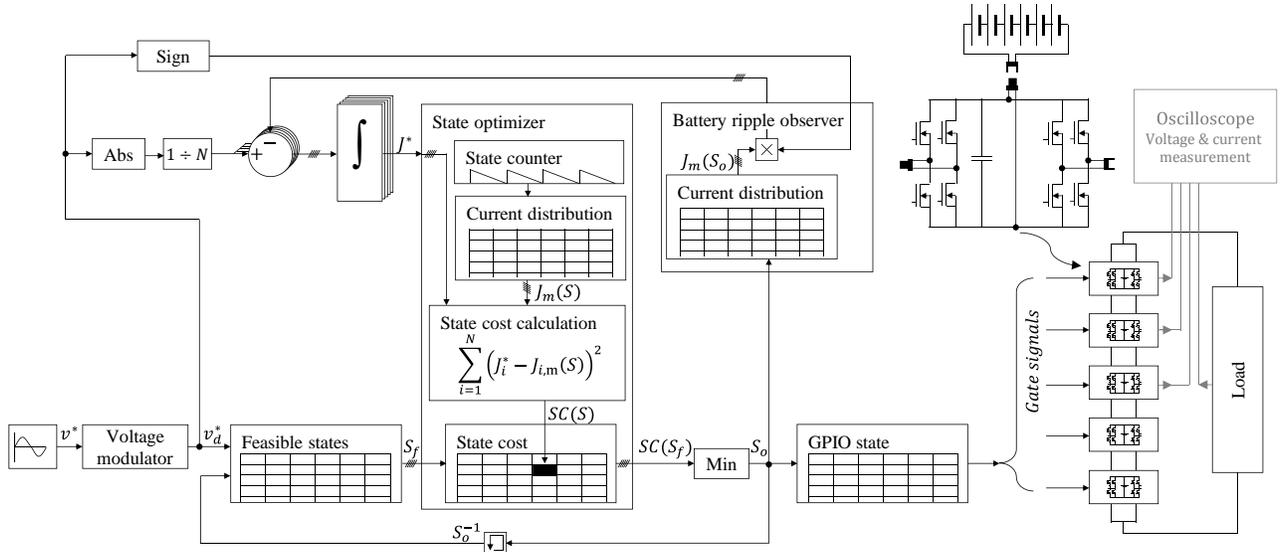

**Figure 4.** Block diagram of the control algorithm adapted for sensorless operation and open-loop control. The ripple-modulator loop works directly with current distribution $J_m$ without the need for scaling it with the phase-current magnitude. The demand of the ripple modulator considers an equal distribution of load and is calculated from the discrete phase-voltage demand divided by the total number of modules, $N$. The ripple modulator uses an integrator to cumulate any disturbances from the demanded load distribution. The cumulated value is used in the optimizer, which evaluates all states according to given criteria and assigns their cost SC in the state cost table. The phase-voltage loop independently selects feasible states according to the demanded voltage and finds the state with minimal cost function, which most effectively compensates cumulated load disturbances. The load distribution of the state is observed, fed back, and cumulated in the ripple modulator.

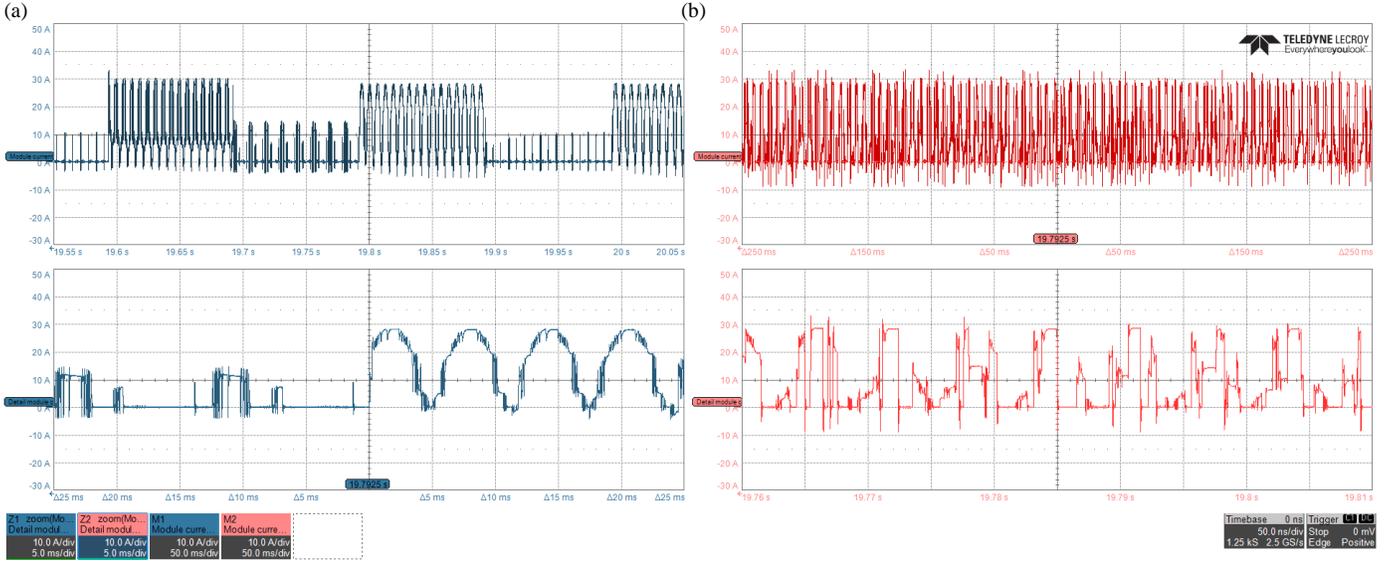

**Figure 6.** Measurement of the current load of an individual module over time of (a) the reference method (with characteristic long-lasting switching patterns) and (b) the proposed method (with distributed switching) measured under phase load specified in Table II. The lower panel of each subplot represents a magnified section out of the full trace.

The implementation of the proposed control method follows the structure given in Figure 4. The method of Specht et al. (see Figure 2) serves as a reference. It delivers measured module load data at an update rate of ~10 Hz with an execution cycle time at 100 ms for the measurement loop, which in turn leads to slacking in the feedback and which we implemented accordingly.

Figure 5 displays the output of the system, while Figure 6 presents the module current comparison between the proposed method with high bandwidth and the reference method from the literature. The slacking feedback of the reference method is noticeable in the module current as it results in artifactual load patterns (see Figure 6a). These artifacts obviously form intervals that correspond to the cycle time of the feedback loop around 100 ms. The output states (and consequently the load distribution) do hardly vary within such an interval, and their repetition generates a regular pattern in the module current provided by the batteries. The patterns cover the entire range from full phase current to practically no battery current in a module at all.

In stark contrast to this established method, the proposed control allows full utilization of the switching rate and evenly distributes the switching events in time. As intended, the distribution of the switching prevents extensive utilization of individual modules and naturally reduces the low frequency content. Yet, the module load can contain some minor residual switching patterns, which we found to be mostly emerging in case of long propagation delay. As long as the period length of the pattern stays negligible compared to the phase frequency of 50 Hz, however, the control method efficiently suppresses low-frequency content through distributed switching (see Figure 6b).

The impact of the proposed control method on the module current is more obvious in the frequency domain. The spectrum of the reference method's module current exhibits dominant peaks at twice the phase load frequency ($f = 2 \times 50$ Hz = 100 Hz). Further peaks can be observer at 5 Hz, which repeats at 10 Hz, 15 Hz, etc. These peaks follow from the update period of 100 ms and the patterns in the module load. The reference method displays relatively low content at frequencies above 100 Hz, which rises again around the switching frequency.

In contrast to the prior art, the frequency content of our proposed method is practically negligible for low frequencies, just starts at 100 Hz, and features increased content up to a fraction of the switching frequency. The reduction of lower frequencies and the partial shift to higher ones is a result of appropriate switching distribution and prevention of pattern formation. Depending on the modulation index, the ripple is shifted to a region corresponding to the individual module switching frequency.

## V. EFFECTIVE IMPEDANCE REDUCTION THROUGH NONFARADAIC SHUNTING

Low-frequency ripple contains larger charge quantities, generates losses, and can age batteries either through the associated heating stress or potentially also electrochemical degradation [48]. Above a certain frequency, however, the impedance of battery cells drops steeper (closer to a dielectrically capacitive $f^{-1}$) than the diffusion-limited $f^{-1/2}$ behavior of faradaic reactions until at very high frequencies the inductance sets a minimum [69]. At such high frequencies, not only the losses decrease but also faradaic processes cease as they cannot follow those charge oscillations anymore; the detected reduction in ageing potential at higher frequencies concurs with this effect [48]. Impedance spectroscopy indicates the transition from faradaic processes, i.e., electrochemical charge-transfer reactions, as the key source of the currents to the dielectric electrode capacitance for most cells above 100 Hz – 1 kHz.

This behavior can be represented with a small-signal approximation of the widely used Randles' equivalent circuit for the electrochemical interface (see inset of Figure 7) [70]. With increasing frequency, the dielectric electrode capacitance bypasses the slow diffusion-limited charge-transfer processes, reducing loss. Consequently, the proposed control method should lead to lower losses.

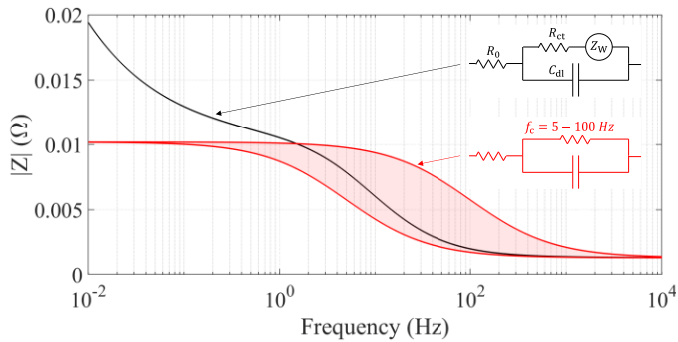

**Figure 7.** Impedance profile over frequency of Randles' model compared to a reduced first-order high-pass filter impedance to represent the dielectric shunting around the diffusion limitation of the faradaic component. Inset: Randles' equivalent circuit with the Warburg impedance $Z_W$ representing the diffusion limitation of the charge-transfer reactions, the charge-transfer equivalent resistance $R_{ct}$ of ion travel near the electrode interface, the double-layer capacitance $C_{dl}$, and the electrolyte/separator equivalent resistance; reduced small-signal filter representation with range of cut-off frequencies $f_c$.

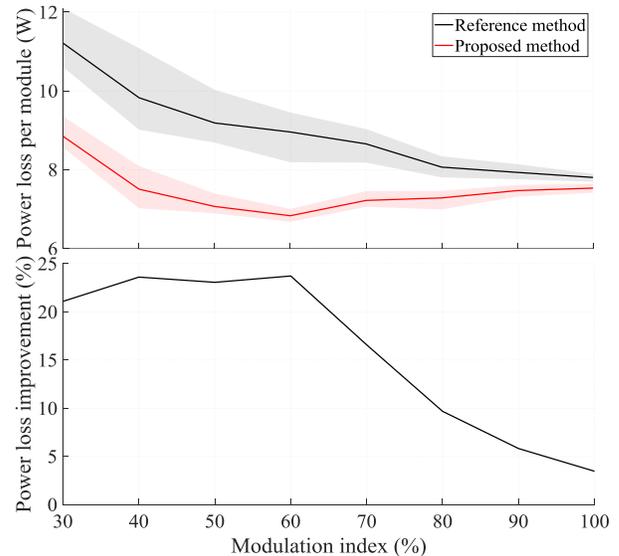

**Figure 8.** Top: Battery power loss averaged across multiple modules and experiments for various modulation indices. The suppression of low-frequency ripple currents and shift of some of those to higher frequencies, where the batteries can absorb the current dielectrically and accordingly show lower effective impedance, clearly reduce the loss and heating in the batteries. Bottom: Power loss improvement of the proposed method with observer relative to reference. The loss improvement gradually decreases on the way to maximum modulation index, where high utilization of modules reduces the available degrees of freedom in circuit reconfiguration for optimization of the ripple

To quantify the losses, we performed load tests of 30 seconds each and extracted the value of the internal voltage $V_i$ shortly after the load test with a delay of 15 seconds for voltage settling to evaluate the internal power loss. We further pool results from multiple measurements and modules to incorporate the entire range of potential conditions.

As intended, the power loss of the proposed scheduling method is throughout all modulation indices substantially lower than the conventional method from the literature as Figure 8 indicates. The solid lines represent the average values, whereas the lighter range indicates the observed variety of values across all measurements. The spread is mainly from module to module, while repetitive measurements within the same module at the same SoC did not vary by more than 0.1 % of the power loss. Our method reduces the power loss by up to 12 W, which equals a reduction in battery loss by ~20 % compared to the reference method. The loss improvement gradually decreases close to 100 % modulation index, where high utilization of modules reduces the degrees of freedom in circuit reconfiguration for optimization of the ripple.

## VI. Conclusion

This paper presents a novel control method for modular multilevel converters with integrated batteries, which suffer from substantial ripple load on the batteries and associated degradation and additional heating. The method aims for battery applications and their specific need for better battery treatment. Based on previous observations that the impedance of the battery cells as well as ageing potential tend to decrease with frequency, since higher frequencies are absorbed by the dielectric electrode capacitance.

In conventional methods, the limited speed and nonnegligible latency of sensor data collection from the individual modules in addition to often slow update rates of scheduling tables typically generate regular patterns in the module load. Accordingly, also module-balancing loops of the state of the art have to adjust their bandwidth to these conditions to avoid instability and driving oscillations. To solve this issue, we developed a battery-ripple observer, which allows us to create a fast feedback loop despite unavoidable latencies so that our control algorithm can actively modulate the module ripple and operate with comparably high bandwidth. We designed a control algorithm to minimize particularly low-frequency content of the module load spectrum.

We evaluated our novel control technique experimentally and compared it to the state of the art. Due to above-mentioned limitations, the conventional method from the literature indeed produced fluctuating module load below 100 Hz and even below 10 Hz, which corresponds to the module data acquisition period, latencies, and bandwidth limits of the feedback, whereas our proposed method could prevent the formation of harmonics in this frequency range and shift fluctuations to the band around 5 kHz. Consequently, our control approach does effectively exploit lowered impedance at higher frequencies and reduces losses in the battery modules by up to 20 %.